\documentclass[12pt]{article}
\usepackage{amsfonts}
\usepackage{amsmath}
\usepackage{theorem}
\usepackage{enumerate}
\usepackage{latexsym}
\usepackage{graphicx}

\def\l{\lambda}

\newtheorem{teo}{Theorem}[section]

\newtheorem{tve}{Proposition}[section]
\newtheorem{nas}{Corollary}[section]

\title{"Doubled'' generalized Landau-Lifshiz  hierarchies
and special quasigraded Lie algebras.}
\author {\sl T. Skrypnyk }
\date{ }
\begin{document}
\thispagestyle{empty}
 \maketitle
  \begin{center}
 {\footnotesize
 Bogoliubov Institute for Theoretical Physics,
 Metrologichna st.14-b,  Kiev 03143, Ukraine.\\
 E-mail: tskrypnyk@imath.kiev.ua\\
 Fax: 380-44-2665998}
 \end{center}
\begin {abstract}
 Using  special quasigraded Lie algebras we obtain new hierarchies of
integrable nonlinear vector equations admitting zero-curvature
representations. Among them the most interesting is extension of
the generalized Landau-Lifshitz hierarchy which we call
"doubled'' generalized Landau-Lifshiz hierarchy. This hierarchy
can be also interpreted as an anisotropic vector generalization of
"modified'' Sine-Gordon hierarchy or as a very special vector
generalization of $so(3)$ anisotropic chiral field hierarchy.
\end{abstract}

\vskip  20 pt

\noindent Short title: "Doubled'' generalized Landau-Lifshiz
hierarchies.

\vskip  10 pt
 \noindent PACS:  02.20.Sv; 02.20 Tw; 02.30.Ik; 02.30.Jr

\newpage
\section{Introduction}
Integrability of equations of $1+1$ field theory and condensed
matter physics is based on the
 the possibility to represent them in the form of the
so-called zero-curvature equations \cite{ZahSh},\cite{TF}:
\begin{equation}\label{z}
\frac{\partial U(x,t,\l)}{\partial t}-\frac{\partial V(x,t,\l)
}{\partial x}+[U(x,t,\l),V(x,t,\l)]=0.
\end{equation}
The most productive interpretation  of zero-curvature equations
 (see \cite{New} and \cite{Hol})
 is to consider them as a consistency
condition for a set of  commuting hamiltonian flows on a dual
space to some infinite-dimensional Lie algebra
$\widetilde{\mathfrak{g}}$ of matrix-valued function of $\l$
written in the Euler-Arnold (generalized Lax) form:
\begin{equation}\label{eae}
\frac{\partial L(\l)}{\partial t_l}=ad^*_{\nabla
I_{l}(L(\l))}L(\l), \ \frac{\partial L(\l)}{\partial
t_k}=ad^*_{\nabla I_{k}(L(\l))}L(\l),
\end{equation}
where $L(\l)\in\widetilde{\mathfrak{g}}^*$ is the generic element
of the dual space, $\nabla I_{k}(L(\l))\in
\widetilde{\mathfrak{g}}$ is the algebra-valued gradient of
$I_{k}(L(\l))$, and the "hamiltonians'' $I_{k}(L(\l))$,
$I_{l}(L(\l))$ belong to the set of mutually commuting with
respect to the natural Lie-Poisson bracket functions on
$\widetilde{\mathfrak{g}}^*$. The consistency condition of two
commuting flows given by equations (\ref{eae})  yields equation
(\ref{z}) with $U \equiv \nabla I_k $, $V \equiv  \nabla I_l $,
 $t_k\equiv x$, $t_l\equiv t$.
In such a way we obtain  a lot of equations in partial
 derivatives that are indexed by two commuting  hamiltonians $I_k$
 and $I_l$. The set of  equations (\ref{z}) with  fixed
 index  $k$ and all indices $l$ constitute so-called "integrable
 hierarchy''. Hence, in
order to construct new integrable hierarchies in the framework of
the described  approach it is necessary to have some
infinite-dimensional Lie algebra $\widetilde{\mathfrak{g}}$
possessing infinite set of mutually commuting hamiltonians
$\{I_k\}$ on its dual space. The main method, that provides such
the set is a famous Kostant-Adler scheme and its extensions
(\cite{RST3},\cite{New}).  Main ingredient of this scheme is an
existence of the decomposition of the algebra
$\widetilde{\mathfrak{g}}$ into sum of two subalgebras:
$\widetilde{\mathfrak{g}}=\widetilde{\mathfrak{g}}_+ +
\widetilde{\mathfrak{g}}_-$.

Although the described above approach was originally  based on
the graded loop algebras $L(\mathfrak{g})=\mathfrak{g}\otimes
P(\l,\l^{-1})$ (\cite{New}, \cite{Hol}) that possess
decompositions into sums of two subalgebras, in the papers
\cite{Hol1}-\cite{Hol3} it was shown that a special Lie algebra
$\mathfrak{g}_{\mathcal{E}}$ on the elliptic curve $\mathcal{E}$
also possess the decomposition
$\mathfrak{g}_{\mathcal{E}}=\mathfrak{g}_{\mathcal{E}}^+
+\mathfrak{g}_{\mathcal{E}}^-$ and could be used in order to
produce integrable systems. In
 our papers
\cite{Skr1}-\cite{Skr3} we have generalized this construction
onto the case of special guasigraded Lie algebras
$\mathfrak{g}_{\mathcal{H}}$ on the algebraic curve
$\mathcal{H}$. With their help we have obtained new integrable
hamiltonian systems (both finite and infinite dimensional)
\cite{Skr2}-\cite{Skr3}. In  papers \cite{Skr4}-\cite{Skr6} we
gave Lie algebraic explanation of our previous semi-geometric
construction of the Lie algebras $\mathfrak{g}_{\mathcal{H}}$.
More explicitly, we have constructed a family of quasigraded Lie
algebras $\mathcal{\mathfrak{g}}_A$  parametrized by some
numerical matrices $A$, such that loop algebras $L(\mathfrak{g})$
correspond to the case $A\equiv 0$ and quasigraded Lie algebras
$\mathfrak{g}_{\mathcal{H}}$ correspond to the case  $A\in
Diag(n)$.

In the present paper we generalize  construction of
\cite{Skr4}-\cite{Skr6}  introducing even larger family of
quasigraded Lie algebras $\widetilde{\mathfrak{g}}_{A_1,A_2}$
numbered by two numerical matrices $A_1$ and $A_2$  to which
Kostant-Adler scheme may be applied. A family of Lie algebras
$\widetilde{\mathfrak{g}}_{A}$ ( see \cite{Skr4}-\cite{Skr6})  is
embedded into the family of Lie algebras
$\widetilde{\mathfrak{g}}_{A_1,A_2}$ as the algebras
$\widetilde{\mathfrak{g}}_{1,A}$.
 We show that three types of integrable hierarchies is  associated  with Lie
algebras $\widetilde{\mathfrak{g}}_{A_1,A_2}$: two small
hierarchies are associated with algebras
$\widetilde{\mathfrak{g}}_{A_1,A_2}^{\pm}$ and large hierarchy is
associated with the Lie algebra
$\widetilde{\mathfrak{g}}_{A_1,A_2}$. We show, that in the case
when both matrices $A_i$ are degenerated, the algebras
$\widetilde{\mathfrak{g}}_{A_1,A_2}$ and
$\widetilde{\mathfrak{g}}_A$ are not isomorphic as quasigraded
Lie algebras. This means that integrable hierarchies associated
with $\widetilde{\mathfrak{g}}_{A_1,A_2}$ such that $\mathrm{det}
A_i=0$  are  not equivalent to the integrable hierarchies
associated with $\widetilde{\mathfrak{g}}_A$ (see
\cite{Skr3},\cite{Skr5}, \cite{Skr6}). Moreover, we show, that
when the matrices $A_i$ have the same matrix rang subalgebras
$\widetilde{\mathfrak{g}}_{A_1,A_2}^{+}$ and
$\widetilde{\mathfrak{g}}_{A_1,A_2}^{-}$ are isomorphic, and
corresponding integrable hierarchies are also equivalent. That is
why "large'' integrable hierarchy associated with the whole Lie
algebra $\widetilde{\mathfrak{g}}_{A_1,A_2}$ could be viewed as
the "double'' of integrable hierarchy associated with
$\widetilde{\mathfrak{g}}_{A_1,A_2}^{\pm}$. The "doubling"
consists in adding of "negative'' flows  and new dynamical
variables to the integrable hierarchy associated with
$\widetilde{\mathfrak{g}}_{A_1,A_2}^{\pm}$.

We consider these hierarchies in the case $\mathfrak{g}=so(n)$
and $\mathrm{rank} A_i=n-1$ in detail.  We show, that integrable
hierarchy associated with $\widetilde{so(n)}^{\pm}_{A_1,A_2}$
coincides with $(n-1)$-component vector generalization of the
ordinary $3$-component Landau-Lifshiz hierarchy.
 For $n>4$ this
hierarchy was first obtained  in $\cite{GS}$ using technique of
"dressing'' and Lie algebra isomorphic to
$\widetilde{\mathfrak{g}}_{1,A}^{+}$ embedded into Lie algebra of
formal power series. Simplest equation of this hierarchy has the
form:
\begin{equation}\label{fi2}
\frac{\partial \overrightarrow{s}}{\partial t}= \frac{\partial
}{\partial x} \Bigl(\dfrac{\partial^2
\overrightarrow{s}}{\partial x^2} + 3/2(\dfrac{\partial
\overrightarrow{s}}{\partial x }, \dfrac{\partial
\overrightarrow{s}}{\partial x })\overrightarrow{s}\Bigr) + 3/2
(\overrightarrow{s},{J}\overrightarrow{s})\dfrac{\partial
\overrightarrow{s} }{\partial x},
\end{equation}
where $\overrightarrow{s}$ is $n-1$-component vector and tensor
of anisotropy $J$ is expressed via $A_1$ and $A_2$.

 "Double'' of the generalized Landau-Lifshiz hierarchy is the
"large'' integrable hierarchy associated with
$\widetilde{so(n)}_{A_1,A_2}$. It is $2(n-1)$-component hierarchy
of vector equations satisfying two additional scalar constraints.
The simplest equation of this hierarchy coincide with the two
$(n-1)$-component vector differential equations of the first
order. We show  that for these equations two scalar constraints
are easily solved and we obtain in the result two non-linear
$(n-2)$-component vector equations of the following form:
\begin{gather}\label{final1}
\partial_{x_+}\overrightarrow{s}_-=\Bigl(c_--(\overrightarrow{s}_-,\overrightarrow{s}_-)\Bigr)^{1/2}
\widehat{J}^{1/2}\overrightarrow{s}_+ ,\\ \label{final2}
\partial_{x_-}\overrightarrow{s}_+=
\Bigl(c_+ -(\overrightarrow{s}_+,\overrightarrow{s}_+)\Bigr)^{1/2}
\widehat{J}^{-1/2}\overrightarrow{s}_-.
\end{gather}
where $\overrightarrow{s}_{\pm}$ are $(n-2)$-component vectors,
$c_{\pm}$ are arbitrary constants
 and  anisotropy matrix
$J$ is connected with matrices $A_i$ in a simple way (see section
(\ref{dllh})).

Equations (\ref{final1}-\ref{final2}) is in a sence a "first
negative equation" or a "first negative flow''of the generalized
L-L hierarchy,  $\overrightarrow{s}_{+}$ is an $n-2$ independent
components of its $n-1$-component vector of dynamical variables:
$\overrightarrow{s}=(s_1,\overrightarrow{s}_{+})$, $x_{+}\equiv x$
is a space coordinate and $x_{-}$ is a first "negative'' time.

 It is necessary
to notice that in the $n=3$  case equations
(\ref{final1}-\ref{final2}) are equivalent to the "modified
Sine-Gordon'' equation \cite{Chen}, \cite{BZ} and in the $n=4$
case to the $so(3)$ anisotropic chiral field equations
\cite{Cher}.

The structure of the present article is the following: in the
second section we introduce algebras
$\widetilde{\mathfrak{g}}_{A_1,A_2}$ and describe their
properties. In the third section we obtain integrable hierarchies
associated with the Lie algebras
$\widetilde{\mathfrak{g}}_{A_1,A_2}$ and its subalgebras
$\widetilde{\mathfrak{g}}_{A_1,A_2}^{\pm}$.
 In the last its subsection  we consider the examples of this construction:
generalized Landau-Lifshiz hierarchy  and  its "double''.

\section{K-A admissible quasigraded Lie algebras. }
\subsection{Lie algebras $\widetilde{\mathfrak{g} }_{A_1,A_2}$.}
 Let $\mathfrak{g}$ be  a classical matrix Lie algebra of the
type $gl(n)$, $so(n)$ and $sp(n)$ over the field of the complex
or real numbers.  We will realize algebra $so(n)$ as algebra of
skew-symmetric matrices: $so(n)=\{X\in gl(n)|X=-X^\top\}$ and
algebra $sp(n)$  as the following  matrix algebra: $sp(n)=\{X\in
gl(n)|X=sX^\top s\}$, where $n$ is an even number, $s\in so(n)$
and $s^2=-1$.

Let us introduce the new Lie bracket into a loop space
 $L(\mathfrak{g})=\mathfrak{g}\otimes
Pol(\l, \l^{-1})$:
\begin{equation}\label{br2}
[X (\l), Y(\l)] = [X(\l),Y(\l)]_{A_1} - \l[X(\l),Y(\l)]_{A_2},
\end{equation}
where $X (\l),Y(\l)\in \mathfrak{g}\otimes Pol(\l,\l^{-1})$,
$A_i$ are the numerical  $n\times n$ matrices,
$[X,Y]_{A_i}=XA_iY-YA_iX$.

Brackets $[X,Y]_{A_i}=XA_iY-YA_iX$ have arisen in the theory of
consistent Poisson brackets on the finite-dimensional Lie
algebras $\mathfrak{g}$ \cite{CP},\cite{Bols}. In the present
paper we use them in order to construct new Lie bracket on the
infinite-dimensional space $\mathfrak{g}\otimes Pol(\l,\l^{-1})$
( see also \cite{Skr4}).

The following proposition holds true:
\begin{tve}
 Let  the numerical  $n\times
n$ matrices $A_i,\ i=1,2$ have the following form:

 1) $A_i$ is arbitrary for $\mathfrak{g}=gl(n)$,

 2) $A_i=A_i^{\top}$ for  $\mathfrak{g}=so(n)$,

 3) $A_i=-sA_i^{\top} s$ for $\mathfrak{g}=sp(n)$.

Then bracket (\ref{br2}) is a correctly defined Lie bracket on
$\mathfrak{g}\otimes Pol(\l,\l^{-1})$.
\end{tve}

{\it Definition.} We will denote infinite-dimensional space
$\mathfrak{g}\otimes Pol(\l,\l^{-1})$ with the Lie bracket given
by (\ref{br2}) by $\widetilde{\mathfrak{g} }_{A_1,A_2}$.

{\bf Remark 1.} Algebra $\widetilde{\mathfrak{g}}_{A_1,A_2}$ could
be realized also in the space of special matrix valued functions
of $\l$ with an ordinary Lie bracket $[\ ,\ ]$. Nevertheless we
consider realization in the space $\mathfrak{g}\otimes
Pol(\l,\l^{-1})$ with the bracket (\ref{br2}) to be the most
convenient.

Now we can introduce the convenient bases in the algebras
$\widetilde{\mathfrak{g} }_{A_1,A_2}$. Due to the fact, that we
are dealing with matrix Lie algebras  $\mathfrak{g}$, we will
denote their basic elements as $X_{ij}$. Let $ X_{ij}^m\equiv
X_{ij}\otimes \l^m$ be the natural basis in
$\widetilde{\mathfrak{g} }_{A_1,A_2}$.  Commutation relations
(\ref{br2}) in this basis  have the following  form:
\begin{equation}\label{bra2}
[X_{ij}^r, X_{kl}^m]=\sum\limits_{p,q}
C_{ij,kl}^{pq}(A_1)X_{pq}^{r+m} - \sum\limits_{p,q}
C_{ij,kl}^{pq}(A_2)X_{pq}^{r+m+1},
\end{equation}
where   $C_{ij,kl}^{pq}(A_i)$ are the structure constants of the
Lie algebras $\mathfrak{g}_{A_i}$.

{\bf Remark 2.} Note that contrary to the case of loop algebras
our algebras $\widetilde{\mathfrak{g} }_{A_1,A_2}$ admit only one
type of decomposition $\widetilde{\mathfrak{g}
}_{A_1,A_2}=\widetilde{\mathfrak{g} }_{A_1,A_2}^+ +
\widetilde{\mathfrak{g} }_{A_1,A_2}^-$ compatible with
quasigrading, where subalgebras $\widetilde{\mathfrak{g}
}_{A_1,A_2}^{\pm}$ are defined in the natural way:
\begin{equation}
\widetilde{\mathfrak{g}
}_{A_1,A_2}^{+}=\mathrm{Span}_{\mathbb{K}}\{X^m_{ij}|m\geq 0\},\
\ \widetilde{\mathfrak{g}
}_{A_1,A_2}^{-}=\mathrm{Span}_{\mathbb{K}}\{X^m_{ij}|m<0 \}.
\end{equation}

Let us now find equivalences among the constructed Lie algebras.
In particular, let us find conditions when
$\widetilde{\mathfrak{g}}_{A_1,A_2}$ is equivalent to the algebra
$\widetilde{\mathfrak{g}}_{A}\equiv\widetilde{\mathfrak{g}}_{1,A}$
introduced in our previous papers \cite{Skr4}-\cite{Skr6}. All
equivalences are understood in the sence of the isomorphisms of
quasigraded Lie algebras. The following Proposition is true:
\begin{tve}
(i) The following isomorphisms hold:
$\widetilde{\mathfrak{g}}^{\pm}_{A_1,A_2}\simeq
\widetilde{\mathfrak{g}}^{\mp}_{A_2,A_1}$,
$\widetilde{\mathfrak{g}}_{A_1,A_2}\simeq
\widetilde{\mathfrak{g}}_{A_2,A_1}$.
\\
(ii) If there exist matrix $C$ such that  $C A_1 C=A_2$ and
$C^2=1$ then $\widetilde{\mathfrak{g}}^{\pm}_{A_1,A_2}\simeq
\widetilde{\mathfrak{g}}^{\mp}_{A_1,A_2}$.
\\
(iii) If $\mathrm{det} A_1\neq 0$ or  $\mathrm{det} A_2\neq 0$
then $\widetilde{\mathfrak{g}}_{A_1,A_2}\simeq
\widetilde{\mathfrak{g}}_{A}$.
\end{tve}

{\bf Remark 3.} Item (iii) of the Proposition means that for the
algebra $\widetilde{\mathfrak{g}}_{A_1,A_2}$ in order not to be
equivalent to the algebra  $\widetilde{\mathfrak{g}}_{A}$ of
\cite{Skr4}, \cite{Skr6} matrices $A_1$ and $A_2$ should be
degenerated. That is why we will consider the case $det A_i =0$
as the main case in the present paper.

\subsection{Coadjoint representation and its invariants.}
In this subsection we  define dual spaces, coadjoint
representations and their invariants for the Lie algebras
$\widetilde{\mathfrak{g}}_{A_1,A_2}$. At first we explicitly
describe the dual space $\widetilde{\mathfrak{g}}_{A_1,A_2}^*$ of
$\widetilde{\mathfrak{g}}_{A_1,A_2}$. For this purpose we define
the pairing between $\widetilde{\mathfrak{g}}_{A_1,A_2}$ and
$\widetilde{\mathfrak{g}}_{A_1,A_2}^*$ in the following way:
\begin{equation}\label{pair}
\langle X,L \rangle=res_{\l=0} Tr(X(\l)L(\l)).
\end{equation}
 The generic element of the dual space
$\widetilde{\mathfrak{g}}_{A_1,A_2}^*$ with respect to this
pairing is written as follows:
\begin{equation}
L(\l)=\sum\limits_{k\in
Z}\sum\limits_{i,j=1,n}l_{ij}^{(k)}\l^{-(k+1)} X_{ij}^*.
\end{equation}

From the  explicit form of the adjoint representation (\ref{br2})
and the pairing (\ref{pair}) it is easy to show that
 the coadjoint action of $\widetilde{\mathfrak{g}}_{A_1,A_2}$ on
$\widetilde{\mathfrak{g}}_{A_1,A_2}^*$ has the  form:
\begin{equation}\label{coadj}
ad_{X(\l)}^* \circ L(\l)=
\mathcal{A}(\l)X(\l)L(\l)-L(\l)X(\l)\mathcal{A}(\l),
\end{equation}
where $X(\l), Y(\l)\in\widetilde{\mathfrak{g}}_{A_1,A_2}$,
$L(\l)\in\widetilde{\mathfrak{g}}_{A_1,A_2}^*$,
$\mathcal{A}(\l)=A_1-\l A_2$.

Having the explicit form of the coadjoint action it is easy to
deduce the next Proposition:
\begin{tve}Let $L(\l)$
be the generic element of $\widetilde{\mathfrak{g}}_{A_1,A_2}^*$.
 Then functions
\begin{equation}\label{inv}
  I^m_{k}(L(\l))= {1}/{m}\ res_{\l=0}
\l^{-(k+1)} Tr(L(\l)\mathcal{A}(\l)^{-1})^m.
\end{equation}
are invariants of the coadjoint representation of the algebra
$\widetilde{\mathfrak{g}}_{A_1,A_2}$.
\end{tve}

{\bf Remark 4.} From the definition of the invariant functions it
follows that  in oder to make algebra
$\widetilde{\mathfrak{g}}_{A_1,A_2}$ satisfy requirement (IR3)
matrix $\mathcal{A}(\l)$  should be nondegenerated. This
condition impose additional requirements on the matrices $A_i$.

\subsection{Lie-Poisson structure.}
Let us introduce Poisson structure in the space
$\widetilde{\mathfrak{g}}_{A_1,A_2}^*$ using the defined above
pairing $\langle \ ,\ \rangle$. It defines Lie-Poisson (
Kirillov-Kostant) bracket on
$P(\widetilde{\mathfrak{g}}_{A_1,A_2}^*)$ in the following
standard way:
\begin{equation}\label{br1}
\{F(L(\l)),G(L(\l))\}=\langle L(\l),[\nabla F(L(\l)),\nabla
G(L(\l))]_{\mathcal{A}(\l)} \rangle,
\end{equation}
where $\nabla F(L(\l))= \sum\limits_{k\in Z}\sum\limits_{i,j=1}^n
\dfrac{\partial F }{\partial l_{ij}^{(k)} }X_{ij}^{k}$, $\nabla
G(L(\l))= \sum\limits_{m\in Z}\sum\limits_{k,l=1}^n
\dfrac{\partial G }{\partial l_{kl}^{(m)} }X_{kl}^{m}$.

 From the Proposition \ref{inv}  and standard considerations
 the next statement follows:
\begin{tve}
Functions $I_{k}^m (L(\l))$ are central for  the Lie-Poisson
bracket (\ref{br1}).
\end{tve}

Let us explicitly calculate Poisson bracket (\ref{br1}).
 It is
easy to show, that for the coordinate functions $l_{ij}^{(m)}$
these brackets will have the following form:
\begin{equation}\label{lpbr}
\{l_{ij}^{(n)},l_{kl}^{(m)}\}=\sum\limits_{p,q}C_{ij,kl}^{pq}(A_1)l_{pq}^{(n+m)}
- \sum\limits_{p,q}C_{ij,kl}^{pq}(A_2)l_{pq}^{(n+m+1)}.
\end{equation}
Lie bracket (\ref{lpbr}) determine the structure of the Lie
algebra isomorphic to $\widetilde{\mathfrak{g}}_{A_1,A_2}$ in the
space of linear functions $\{l^{n}_{ij}\}$. That is why the
corresponding Poisson algebra possess decomposition into direct
sum of two Poisson subalgebras or, by other words subspaces
$(\widetilde{\mathfrak{g}}^{\pm}_{A_1,A_2})^*$ are Poisson.

\section{Integrable hierarchies associated with algebras $\widetilde{\mathfrak{g}}_{A_1,A_2}$}
 In this section we  construct two infinite sets of mutually
Poisson-commuting functions on the Lie algebra
$\widetilde{\mathfrak{g}}_{A_1,A_2}$ and Lax type representation
for the corresponding hamiltonian equations. We also derive
zero-curvature equations as a compatibility condition of the
above commuting hamiltonian  flows and consider examples of the
equations in partial derivatives from the corresponding integrable
hierarchies.
\subsection{Infinite-component hamiltonian systems on
$\widetilde{\mathfrak{g}}_{A_1,A_2}^*$.} In this subsection we
construct hamiltonian systems on the infinite-dimensional space
$\widetilde{\mathfrak{g}}_{A_1,A_2}^{*}$ possessing infinite
number of  independent, mutually commuting integrals of motion.

 Let $L^{\pm}(\l)$ be the
generic element of the space
$\widetilde{\mathfrak{g}}_{A_1,A_2}^{\mp*}$:
$$L^{\pm}(\l)\equiv \sum\limits_{i,j=1,n}L^{\pm}_{ij}(\l)X_{ji}
=\sum\limits_{k \in \mathbb{Z}_{\mp}}
\sum\limits_{i,j=1,n}l_{ij}^{(k)}\l^{-(k+1)}X_{ji}.$$
 Let us consider the
restriction of the  invariant functions $\{I^m_{k}(L(\l)) \}$ onto
these subspaces.  Note, that although  Poisson subspaces
$\widetilde{\mathfrak{g}}^{\mp*}_{A_1,A_2}$ are
infinite-dimensional,  functions $\{I^m_{k}(L^{\pm}(\l))\}$ are
polynomials, i.e. after the restriction onto
$\widetilde{\mathfrak{g}}^{\mp*}_{A_1,A_2}$ no infinite sums
appear in their explicit expressions. Let us now  consider
functions $I^m_{k}(L^{\pm}(\l))$ as functions on the whole space
$\widetilde{\mathfrak{g}}_{A_1,A_2}^*$. We  have two sets of
hamiltonians $\{I^{m+}_k(L(\l))\}$ and $\{I^{m-}_k(L(\l))\}$ on
$\widetilde{\mathfrak{g}}_{A_1,A_2}^*$ defined as follows:
$$I^{m\pm}_k(L(\l))\equiv I^{m}_k(L^{\pm}(\l)).$$
 Hamiltonian  flows
corresponding to hamiltonians $I^{m\pm}_k(L(\l))$ are written in
a standard way:
\begin{equation}\label{ham}
\frac{\partial L_{ij}(\l)}{\partial
t^{m\pm}_k}=\{L_{ij}(\l),I^{m\pm}_k(L(\l))\}.
\end{equation}
  The following theorem is true:
\begin{teo}\label{scomm} (i) Hamiltonian equations (\ref{ham}) are
written in the generalized Lax form:
\begin{equation}\label{dlax2}
 \frac{\partial L(\l)}{\partial t^{m\pm}_k}=ad^*_{V^{m\pm}_k(\l)}L(\l)=\mathcal{A}(\l)V^{m\pm}_k(\l)L(\l)-
L(\l)V^{m\pm}_k(\l)\mathcal{A}(\l).
\end{equation}
where $V^{m\pm}_k(\l)=\nabla I^{m\pm}_k(L(\l))\equiv\sum\limits_{s
\in \mathbb{Z}_{\pm}}\sum\limits_{i,j=1}^n \dfrac{\partial
I^{m\pm}_k}{\partial l_{ij}^{(s)} }X_{ij}^{s}$.
\\

(ii)The functions $\{I^{m\pm}_k(L(\l))\}$ form the commutative
subalgebra in the algebra of polynomial functions on
$\widetilde{\mathfrak{g}}_{A_1,A_2}^*$:
$\{I^{m\pm}_k(L(\l)),I^{n\pm}_l(L(\l))\}=
\{I^{m\mp}_k(L(\l)),I^{n\pm}_l(L(\l))\}=0,$ i.e. time flows
defined by equations (\ref{ham})( or (\ref{dlax2}) ) mutually
commute.

(iii) The functions $I^{n\pm}_{l}(L(\l))$ are constant along all
time flows : $\dfrac{\partial I^{n\pm}_{l}}{\partial t^{m\pm}_k}=
\dfrac{\partial I^{n\pm}_{l}}{\partial t^{m\mp}_k}=0.$
\end{teo}
The proof of this theorem repeats the proof of the analogous
theorem for the case of ordinary loop algebras (see \cite{New} and
references therein).

{\bf Remark 5. } Due to the fact that the subspaces
$(\widetilde{\mathfrak{g}}^{\mp}_{A_1,A_2})^*$ are Poisson
equations (\ref{ham}), generated by hamiltonians
$I^{m\pm}_k(L(\l))$, could be restricted onto them, i.e. it is
correctly to consider the following hamiltonian equations:
\begin{equation}\label{hampm}
\frac{\partial L^{+}_{ij}(\l)}{\partial
t^{m+}_k}=\{L^{+}_{ij}(\l),I^{m}_k(L^+(\l))\},\ \ \ \frac{\partial
L^{-}_{ij}(\l)}{\partial
t^{m-}_k}=\{L^{-}_{ij}(\l),I^{m}_k(L^-(\l))\}.
\end{equation}
In particular the following Corollary of the Theorem \ref{scomm}
holds true:
\begin{nas}\label{fcomm} (i) Hamiltonian equations (\ref{hampm}) are
written in the generalized Lax form:
\begin{equation}\label{dlax1}
 \frac{\partial L^{\pm}(\l)}{\partial t^{m\pm}_k}=
 \mathcal{A}(\l)V^{m\pm}_k(\l)L^{\pm}(\l)-
 L^{\pm}(\l)V^{m\pm}_k(\l)\mathcal{A}(\l),
\end{equation}
where $V^{m\pm}_k(\l)$ is defined as in theorem \ref{scomm}.\\

(ii) The functions $\{I^{m}_k(L^{\pm}(\l))\}$ form commutative
subalgebra in the algebra of polynomial functions on
$(\widetilde{\mathfrak{g}}_{A_1,A_2}^{\mp})^*$ and corresponding
time flows mutually commute.

(iii)The functions $\{I^{m}_k(L^{\pm}(\l))\}$ are constant along
time flows (\ref{hampm}).
\end{nas}

In other words Theorem \ref{scomm} provides us with three types of
infinite-component hamiltonian systems possessing infinite sets
of commuting integrals of motion: two "small'' hamiltonian
subsystems on the subspaces
$(\widetilde{\mathfrak{g}}^{\mp}_{A_1,A_2})^*$ with the sets of
integrals $\{I^{n\pm}_{l}\}$ and the "large'' hamiltonian system
on $\widetilde{\mathfrak{g}}^{*}_{A_1,A_2}$ with both sets of
integrals $\{I^{n-}_{l}\}$ and $\{I^{m+}_{k}\}$.

 From the
Theorem \ref{scomm} also follows that due to the commutativity of
time flows  it makes sense to consider $L(\l)$ as functions of
independent time variables $t^{m+}_k$ and $t^{n-}_l$ and consider
simultaneously all equations (\ref{dlax2}) as a system of
differential identities of the first order on all coordinate
functions $l_{ij}^{(s)}(t^{m+}_k, t^{n-}_l)$ that holds true on
the "Liuville torus'' which is the level set of the integrals of
motion: $\{ I^n_{l}(L^{+})=c^{n+}_l, I^m_{k}(L^{-})=c^{m-}_k$.
From this system of differential identities  of the first order
one can extract some finite subsystems of differential
identities  on some finite subsets of   the coordinate functions
$l_{ij}^{(s)}$. These identities are the wanted {\it integrable
equations in the partial derivatives}. In order to have a
systematic procedure  of obtaining of such equations from the
equations (\ref{dlax2}), it is better to consider  equivalent
system of equations instead of (\ref{dlax2}). These will be
zero-curvature equations with the values in
$\widetilde{\mathfrak{g}}_{A_1,A_2}$.
\subsection{Zero-curvature conditions associated with algebras
$\widetilde{\mathfrak{g}}_{A_1,A_2}$.} Considering consistency
conditions of the commuting Lax-type equations (\ref{dlax2}) it
is possible to prove the following theorem:
\begin{teo}\label{dzc} Let infinite-dimensional Lie algebras
$\widetilde{\mathfrak{g}}_{A_1,A_2}$,
 $\widetilde{\mathfrak{g}}_{A_1,A_2}^{\pm}$, their dual spaces
   and polynomial hamiltonians $I_k^m(L^{\pm}(\l))$,
   $I^n_l(L^{\pm}(\l))$ on them be defined as in previous sections.
    Then system of consistent generalized Lax equations
    (\ref{dlax2}) is equivalent to the system of
"deformed'' zero-curvature equations:
\begin{equation}\label{dzce1}
\dfrac{\partial \nabla I_k^m(L^{\pm}(\l))}{\partial
t^{n\pm}_l}-\dfrac{\partial \nabla I^n_l(L^{\pm}(\l))}{\partial
t^{m\pm}_k}+[\nabla I_k^m(L^{\pm}(\l)),\nabla
I^n_l(L^{\pm}(\l))]_{\mathcal{A}(\l)}=0,
\end{equation}
\begin{equation}\label{dzce2}
\dfrac{\partial \nabla I_k^m(L^{\pm}(\l))}{\partial
t^{n\mp}_l}-\dfrac{\partial \nabla I^n_l(L^{\mp}(\l))}{\partial
t^{m\pm}_k}+[\nabla I_k^m(L^{\pm}(\l)),\nabla
I^n_l(L^{\mp}(\l))]_{\mathcal{A}(\l)}=0.
\end{equation}
\end{teo}
(Proof of this theorem repeats proof of the analogous theorem
 for algebras $\widetilde{\mathfrak{g}}_{\mathcal{H}}$  (see
\cite{Skr3}).)

{\bf Remark 6.} Using mentioned above realizations of
$\widetilde{\mathfrak{g}}_A$
 "deformed'' zero-curvature equations can be rewritten
in the form of the standard zero-curvature equations, but in this
case corresponding $U-V$ pairs will be more complicated and we
will work with zero-curvature equations in the "deformed'' form
(\ref{dzce1})-(\ref{dzce2}).

The  theorem above give us possibility to distinguish
 three types of hierarchies, connected
with algebras $\widetilde{\mathfrak{g}}^{\pm}_{A_1,A_2}$ and
$\widetilde{\mathfrak{g}}_{A_1,A_2}$. Integrable hierarchies
associated with subalgebras
$\widetilde{\mathfrak{g}}^{\pm}_{A_1,A_2}$ are described only by
equations (\ref{dzce1}), while integrable hierarchies associated
with $\widetilde{\mathfrak{g}}_{A_1,A_2}$ are described by  both
equations (\ref{dzce1}) and  (\ref{dzce2}),  reflecting the fact,
that we have in this case both positive and negative flows. In
other words, integrable hierarchies associated with the algebras
$\widetilde{\mathfrak{g}}^{\pm}_{A_1,A_2}$ could be viewed as
subhierarchies of the integrable hierarchy associated with
algebra  $\widetilde{\mathfrak{g}}_{A_1,A_2}$. Nevertheless they
are completely self-contained and could be considered separately.
In particular, they do not depend on the "large'' algebra in which
we embed corresponding  subalgebra
$\widetilde{\mathfrak{g}}^{+}_{A_1,A_2}$ or
$\widetilde{\mathfrak{g}}^{-}_{A_1,A_2}$.

 For the case of integrable hierarchies, associated
with algebras $\widetilde{\mathfrak{g}}^{\pm}_{A_1,A_2}$, a choice
of the one of the matrix gradients $\nabla I^{m}_k$ to be
$U$-operator yields fixation of dynamical variables that coincide
with its matrix elements. For the case of integrable systems,
connected with the algebras $\widetilde{\mathfrak{g}}_{A_1,A_2}$
there are two types of hamiltonians and two types of flows. That
is why in this case number of independent dynamical variables may
be doubled: their role is played by the matrix elements of two $U$
operators: $U_+=\nabla I^{m}_k(L^{+}(\l))$ and $U_-=\nabla
I^{n}_l(L^{-}(\l))$, where hamiltonians $I^{m}_k(L^{+}(\l))$ and
$I^{n}_l(L^{-}(\l))$ generate evolution with respect to "times''
$x_+$ and  $x_-$
--- "space'' flows of the hierarchies associated with subalgebras
$\widetilde{\mathfrak{g}}_{A_1,A_2}^{\pm}$.

 The number of the dynamical variables for the chosen
integrable hierarchy coincide with the number of independent
matrix elements of the $U_{\pm}$-operators, where $U_+=\nabla
I^{m}_k(L^{+}(\l))$ and $U_-=\nabla I^{n}_l(L^{-}(\l))$. In our
case, when $I_k^m(L^{\pm}(\l))$ depends on the additional
parameters (matrix elements of the matrices $A_i$) we may
 decrease the number of the dynamical variables
manipulating by these parameters (in particular tending some of
them to zero). Hence, this provides us with simple procedure  of
reduction of the number of functional degrees of freedom.
 We will illustrate this
  in the next subsection on the $\mathfrak{g}=so(n)$ example.

\subsection{Integrable subhierarchy associated  with subalgebra
$\widetilde{so(n)}_{A_1,A_2}^{\pm}$.} The aim of this subsection
is a derivation of the equations of integrable hierarchy connected
with the algebra $\widetilde{\mathfrak{g}}_{A_1,A_2}^{+}$, where
$\mathfrak{g}=so(n)$, matrices $A_i$ are degenerated:
$\mathrm{det} A_i=0$ and $\mathrm{rank}\ A_i=n-1$.
 We
will start our consideration with nondegenerated  case:
$\mathrm{rank}\ A_i=n$ and obtain the case $\mathrm{rank}\
A_i=n-1$ as its continuous limit.

Let us now illustrate the  procedure of obtaining of integrable
equations in the partial derivatives starting from the Lie
algebras $\widetilde{\mathfrak{g}}_{A_1,A_2}^{+}$ where
$\mathfrak{g}$ and $A_i$ are as the described above. For this
purpose we have to describe the set of commuting integrals on
$(\widetilde{so(n)}_{A_1,A_2}^{+})^*$. Let us at first note, that
generic element of the dual space
$(\widetilde{so(n)}_{A_1,A_2}^{+})^*$ has the following form:
\begin{equation}
L^{-}(\l) =\l^{-1}L^{(0)} +\l^{-2} L^{(1)} +\l^{-3} L^{(2)}
+\l^{-4} L^{(3)}+ \cdots,
\end{equation}
 where $L^{(k)}\equiv
\sum\limits_{i<j=1,n}l_{ij}^{( k)}X_{ji}$. We will be interested
in the second order integrals (hamiltonians).  By the very
definition they are written as follows:
\begin{equation}\label{soint}
  I^{2-}_{k}(L(\l))=1/2\ res_{\l=0}
\l^{-(k+1)} Tr(L^{-}(\l)\mathcal{A}(\l)^{-1})^2.
\end{equation}
In order for hamiltonians $I^{2-}_k$ to be polynomials we will use
the decomposition of the matrix $\mathcal{A}(\l)^{-1}$ in the
formal power series in a neighbourhood of infinity:
\begin{gather}\label{soi2}
    I^{2-}(L(\l))=
 Tr\bigl( (\l^{-1}L^{(0)} +\l^{-2} L^{(1)}+
\cdots)(1+A_1A_2^{-1}\l^{-1} +\cdots)A_2^{-1}\l^{-1}\bigr)^2.
\end{gather}
Commuting integrals of the series $I^2(L^-(\l))$ contain
expression $ {A}_2^{-1}$ and in the limit $\mathrm{det} {A}_2=0$
should be regularized in the appropriate way. We will calculate
these hamiltonians for the case $\mathrm{det} A_2 \neq 0$  and
then consider the continuous limit $\mathrm{det} A_2\rightarrow
0$.

Simplest hamiltonians  of the set  (\ref{soi2}) are the functions
$I^{2-}_{-4}(L(\l))$ and $I^{2-}_{-5}(L(\l))$ \footnote{In the
case of the nondegenerated matrices $A_i$ corresponding matrix
gradients produce "integrable anisotropic deformation'' of the
generalized Heisenberg magnet equations \cite{Skr3}.}:
\begin{equation}
 I^{2-}_{-4}(L(\l))=
1/2 Tr (A_2^{-1}L^{(0)})^2,\  \ I^{2-}_{-5}(L(\l))= Tr
(A_2^{-1}L^{(0)}A_1A_2^{-2}L^{(0)})+
(A_2^{-1}L^{(1)}A_2^{-1}L^{(0)}).
\end{equation}
We will hereafter consider the case of the diagonal matrices
$A_i$: $A_1=diag(a^{(1)}_1,a^{(1)}_2,...,a^{(1)}_n),$ $A_2
=diag(a^{(2)}_1,a^{(2)}_2,...,a^{(2)}_{n-1},a^{(2)}_n)$ and take
in the previous formulas continuous limit $a_n^{(2)}\rightarrow
0$. Due to the fact that hamiltonians $I^{2-}_{-4}(L(\l))$ and
$I^{2-}_{-5}(L(\l))$ are singular in this limit we have to
rescale them,  considering the limit $a_n^{(2)}\rightarrow 0$ of
commuting integrals
$$I^{2-'}_{-4}(L(\l))\equiv a_n^{(2)}I^{2-}_{-4}(L(\l)),\ \
I^{2-'}_{-5}(L(\l))\equiv\bigl((a^{(2)}_n/a^{(1)}_n)I^2_{-5}(L^-(\l))
-I^2_{-4}(L^-(\l))\bigr).$$
 Taking this limit we obtain:
\begin{gather}
 I^{2-'}_{-4}(L(\l))=
\sum\limits_{i<n}\dfrac{(l^{(0)}_{in})^2}{a_i^{(2)}},\ \
I^{2-'}_{-5}(L(\l))=\frac{1}{a^{(1)}_n}\sum\limits_{i<n}(2\dfrac{l^{(1)}_{in}l^{(0)}_{in}}{a^{(2)}_i}+
\dfrac{(l^{(0)}_{in})^2a^{(1)}_i}{(a^{(2)}_i)^2}) -
\sum\limits_{0<i<j<n}\dfrac{(l^{(0)}_{ij})^2}{a^{(2)}_ia^{(2)}_j}.
\end{gather}
Corresponding matrix gradients are written as follows:
\begin{equation}\label{gra}
 1/2 \nabla I^{2-'}_{-4}=\sum\limits_{i<n}
 \dfrac{l^{(0)}_{in}}{a^{(2)}_i}X_{in} ,
\  \
 1/2\nabla I^{2-'}_{-5}= \frac{1}{a^{(1)}_n}
\sum\limits_{i<n}\bigl(\l\dfrac{l^{(0)}_{in}}{a^{(2)}_i}
X_{in}+(\dfrac{l^{(1)}_{in}}{a^{(2)}_i}
+\dfrac{a^{(1)}_il^{(0)}_{in}}{(a^{(2)}_i)^2}) X_{in}\bigr)
 -
\sum\limits_{i<j<n}\dfrac{l^{(0)}_{ij}}{a^{(2)}_ia^{(2)}_j}X_{ij}.
\end{equation}

Let us take for the hamiltonian that generate space flow function
$I^{2-'}_{-4}$. This fix our integrable hierarchy with
$U\equiv\nabla I^{2-'}_{-4}$. Taking into account the explicit
form of  $\nabla I^{2-'}_{-4}$ we obtain that dynamical variables
in the corresponding hierarchy are functions $l^{(0)}_{in}$,
$i\in 1,n-1$. Hence, by taking the limit $a^{(2)}_n\rightarrow
0$,  we have decreased the number of functional degrees of
freedom from $n(n-1)/2$ (number of the independent components of
$\nabla I^{2-}_{-4}$) to $n-1$ (number of the independent
components of $\nabla I^{2-'}_{-4}$).

In order to obtain all equations of this hierarchy it is necessary
to obtain the regularized expression $\nabla I^{m-'}_{k}$ for the
all other hamiltonians $\nabla I^{m-}_{k}$ to  express  all
coordinate functions $l^{(k)}_{ij}$ $k\geq 0$ via $l^{(0)}_{in}$
and its derivatives with respect to the coordinate $x$ and
substitute them into expression for $\nabla I^{m-'}_{k}$ into
zero-curvature equation:
\begin{equation}\label{zce3}
\dfrac{\partial \nabla I^{2-'}_{-4}}{\partial
t^{m'}_k}=\dfrac{\partial \nabla I^{m-'}_k}{\partial x}-[\nabla
I^{2-'}_{-4},\nabla I^{m-'}_k]_{A(\l)}.
\end{equation}
We will consider the simplest  equation of the hierarchy
(\ref{zce3}) that correspond to the time flow of the hamiltonian
$I^{2-'}_{-5}$, i.e. we will put $V\equiv \nabla I^{2'}_{-5}$.
Coordinate functions $l^{(0)}_{ij}$  and $l^{(1)}_{in}$ where
$i,j\in 1,n-1$ enter in the explicit expression of the
$V$-operator (\ref{gra}). They should be expressed via
$l^{(0)}_{in}$ and their derivatives in order to obtain wanted
equation on the dynamical variables $l^{(0)}_{in}$. This can be
achieved by decomposing both sides of equation (\ref{zce3})  in
the powers of spectral parameter $\l$. Rescaling  time variables
$x\rightarrow 2x$, $t\rightarrow 2t$ and introducing the
following notations: $ m^{(1)}_i= l^{(1)}_{in}
+\dfrac{a^{(1)}_il^{(0)}_{in}}{a^{(2)}_i},$ we  obtain that for
the chosen $U-V$ pair  zero-curvature equation (\ref{zce3}) is
equivalent to the following system of differential equations:
\begin{equation}\label{one}
\frac{\partial l^{(0)}_{in}}{\partial t}-\frac{\partial
m^{(1)}_i}{\partial
x}=\sum\limits_{k=1}^{n-1}\dfrac{l^{(0)}_{ik}a^{(1)}_kl^{(0)}_{kn}}{(a^{(2)}_k)^2},
\end{equation}
\begin{equation} \label{two}
\frac{\partial l^{(0)}_{in}}{\partial
x}=\sum\limits_{k=1}^{n-1}\frac{l^{(0)}_{ik}l^{(0)}_{kn}}{a^{(2)}_k},
\end{equation}
\begin{equation} \label{three}
\frac{\partial  l^{(0)}_{ij}}{\partial x
}=a^{(1)}_n(m^{(1)}_{i}l^{(0)}_{jn}-m^{(1)}_{j}l^{(0)}_{in}).
\end{equation}
We will use equations (\ref{two}) and (\ref{three})  in order to
 to express $m^{(1)}_i$ and $l^{(0)}_{ij}$ via dynamical variables $l^{(0)}_{jn}$
and their $x$-derivatives. From these  equations
 it is easy to deduce, that the
following equalities hold true:
\begin{gather}\label{expres1}
l^{(0)}_{ij}=\dfrac{\partial l^{(0)}_{in}}{\partial x}
l^{(0)}_{jn}-\dfrac{\partial l^{(0)}_{jn}}{\partial x}
l^{(0)}_{in} ,\\ \label{expres2} m^{(1)}_i=
1/a^{(1)}_n\dfrac{\partial^2 l^{(0)}_{in}}{\partial x^2}
+c_2(L^{(0)})l^{(0)}_{jn},
\end{gather}
 where $c_2(L^{(0)})$
is some scalar function of the dynamical variables $l^{(0)}_{in}$.
We  determine explicit form of the function $c_2(L^{(0)})$ using
the fact  that hamiltonians $I^{2'}_{-4}$ and $I^{2'}_{-5}$ are
constant along all flows and we may put $I^{2'}_{-4}=1$,
$I^{2'}_{-5}=0$. Using this and
 introducing vector
 $(\overrightarrow{l})_i=l^{(0)}_{in}/a^{(2)}_i$ and matrices $A_i'\in Mat(n-1)$, $A_i'
=diag(a^{(i)}_1,a^{(i)}_2,a^{(1)}_3,...,a^{(i)}_{n-1})$ we obtain:
$$
c_2(\overrightarrow{l})=1/2(\overrightarrow{l},A_1'\overrightarrow{l})
+ 1/a^{(1)}_n\cdot 3/2 (\dfrac{\partial
\overrightarrow{l}}{\partial x }, A_2'\dfrac{\partial
\overrightarrow{l}}{\partial x }).$$ Using equality
(\ref{expres1}) we also deduce that:
$$\sum\limits_{k=1}^{n-1}\dfrac{l^{(0)}_{ik}a^{(1)}_kl^{(0)}_{kn}}{a^{(2)}_i(a^{(2)}_k)^2},
=-
1/2\dfrac{\partial(\overrightarrow{l},A_1'\overrightarrow{l})}{\partial
x}(\overrightarrow{l})_i
+(\overrightarrow{l},A_1'\overrightarrow{l})\dfrac{\partial
(\overrightarrow{l})_i }{\partial x}.$$ In the result we obtain
the following differential equation in the partial derivatives:
\begin{equation}\label{fi1}
\frac{\partial \overrightarrow{l}}{\partial
t}=\frac{1}{a_n^{(1)}}\frac{\partial }{\partial x}
\Bigl(\dfrac{\partial^2 \overrightarrow{l}}{\partial x^2} +
3/2(\dfrac{\partial \overrightarrow{l}}{\partial x },
A_2'\dfrac{\partial \overrightarrow{l}}{\partial x
})\overrightarrow{l}\Bigr) + 3/2
(\overrightarrow{l},A_1'\overrightarrow{l})\dfrac{\partial
\overrightarrow{l} }{\partial x}.
\end{equation}
In order to transform this equation to a more standard form it is
necessary to introduce new notations:
$\overrightarrow{s}=(A_2')^{1/2} \overrightarrow{l}$, ${J}\equiv
A_1'(A_2')^{-1}$. Under such a replacement of variables
 constraint
$(\overrightarrow{l},A_2'\overrightarrow{l})=1$ pass to the
standard constraint $(\overrightarrow{s},\overrightarrow{s})=1$
and equation (\ref{fi1}) to the  higher Landau-Lifshiz equation:
\begin{equation}\label{fi21}
\frac{\partial \overrightarrow{s}}{\partial t}=\frac{1}{a_n^{(1)}}
\frac{\partial }{\partial x} \Bigl(\dfrac{\partial^2
\overrightarrow{s}}{\partial x^2} + 3/2(\dfrac{\partial
\overrightarrow{s}}{\partial x }, \dfrac{\partial
\overrightarrow{s}}{\partial x })\overrightarrow{s}\Bigr) + 3/2
(\overrightarrow{s},{J}\overrightarrow{s})\dfrac{\partial
\overrightarrow{s} }{\partial x}.
\end{equation}

{\bf Remark 7.} In the case $n=4$ this equation is the higher
equation of the Landau-Lifshiz hierarchy.  For $n>4$ this
equation was first obtained  in \cite{GS} using the technique of
"dressing'' and the embedding of the( specially realized )
algebra $\widetilde{so(n)}^+_{A}$  into  algebra $so(n)((\l))$ of
formal power series. Equation (\ref{fi21}) was  also obtained in
our previous paper \cite{Skr6} using the algebra
$\widetilde{so(n)}^+_{A}$ naturally embedded into algebra
$\widetilde{so(n)}_{A}$.

In the next subsection we will obtain the simplest equation of
 "doubled'' Landau-Lifshiz hierarchy. In order to do
so it is necessary to use not the algebra of formal power series,
nor the algebra $\widetilde{so(n)}_{A}$ but its generalization
--- Lie algebra $\widetilde{so(n)}_{A_1,A_2}$.

\subsection{Integrable hierarchy associated  with algebra
$\widetilde{so(n)}_{A_1,A_2}$.}\label{dllh}
 In this subsection
we will consider  integrable hierarchies, admitting zero
curvature type representation with $U-V$ pairs taking  values in
the algebra $\widetilde{so(n)}_{A_1,A_2}$. In order to obtain
"double'' of Landau-Lifshiz hierarchies  it is necessary to
consider the case of  the matrix $\mathcal{A}(\l)$ formed by the
degenerated matrices $A_i$, such that  $\mathrm{rank} A_i=n-1$
but $\mathrm{rank} \mathcal{A}(\l)=n$. As in the previous example
of the hierarchies connected with
$\widetilde{so(n)}_{A_1,A_2}^{\pm}$ we will at first consider the
case of the nondegenerated matrices $A_i$: $\mathrm{rank}\ A_i=n$
and obtain the case $\mathrm{rank}\ A_i=n-1$ as its continuous
limit.

 Let us now illustrate the  procedure of obtaining integrable
equations in the partial derivatives associated with algebras
$\widetilde{so(n)}_{A_1,A_2}$. For this purpose we have to
describe the set of commuting integrals on
$(\widetilde{so(n)}_{A_1,A_2})^*$. Generic elements of the dual
spaces to subalgebras $(\widetilde{so(n)}_{A_1,A_2}^{-})^*$ and
$(\widetilde{so(n)}_{A_1,A_2}^{+})^*$ have the following form:
\begin{equation}
L^{+}(\l) =L^{(-1)} +\l L^{(-2)} +\l^2 L^{(-3)} + \l^3 L^{(-4)}
+\cdots,
\end{equation}
\begin{equation}
L^{-}(\l) =\l^{-1}L^{(0)} +\l^{-2} L^{(1)} +\l^{-3} L^{(2)}
+\l^{-4} L^{(3)}+ \cdots,
\end{equation}
 where $L^{(\pm k)}\equiv
\sum\limits_{i<j=1,n}l_{ij}^{(\pm k)}X_{ji}$. Second order
integrals (hamiltonians)  by the very definition are written as
follows:
\begin{equation}\label{soint1}
  I^{2\pm}_{k}(L(\l))=1/2\ res_{\l=0}
\l^{-(k+1)} Tr(L^{\pm}(\l)\mathcal{A}(\l)^{-1})^2.
\end{equation}
Let us at first consider hamiltonians $I^{2\pm}_k$ in the case of
the nondegenerated matrices $A$. In order for hamiltonians
$I^{2\pm}_k$ to be polynomials we will use two different
decompositions of the matrix $\mathcal{A}(\l)^{-1}$ in the formal
power series --- in the neighborhood of zero and infinity.
Corresponding hamiltonians are calculated using their own
decompositions:
\begin{gather}\label{soint2}
  I^{2+}_k(L(\l))=
 1/2\ res_{\l=0} \l^{-(k+1)} Tr\bigl(A_1^{-1}(1+A_1^{-1}A_2\l +\cdots
)(L^{(-1)} +\l
L^{(-2)} +\cdots)\bigr)^2,\\
  I^{2-}_k(L(\l))=
 1/2\ res_{\l=0} \l^{-(k+1)}Tr\bigl( (1+A_1A_2^{-1}\l^{-1}
+\cdots)A_2^{-1}\l^{-1}(\l^{-1}L^{(0)} +\l^{-2} L^{(1)}+
\cdots)\bigr)^2.
\end{gather}
Simplest hamiltonians  of these sets  are functions
$I^{2-}_{-4}(L(\l))$ and   $I^{2+}_{0}(L(\l))$ \footnote{In the
case of the nondegenerated matrices $A_i$ corresponding matrix
gradients produce anisotropic chiral field-type equations
\cite{Skr5}.}:
\begin{equation}
 I^{2-}_{-4}(L(\l))=
1/2 Tr (A_2^{-1}L^{(0)})^2, \ \ \  I^{2+}_{0}(L(\l))= 1/2 Tr
(A_1^{-1}L^{(-1)})^2.
\end{equation}
Without loss of generality we will put that  matrices $A_i$ are
diagonal: $A_1=\mathrm{diag}(a^{(1)}_1,...,a^{(1)}_n)$,
$A_2=\mathrm{diag}(a^{(2)}_1,...,a^{(2)}_n)$ and consider the
limits $a_1^{(1)}\rightarrow 0$, $a_n^{(2)}\rightarrow 0$ that
correspond to the simplest degeneration of the matrices $A_i$.
Due to the fact that hamiltonians $I^{2-}_{-4}$ and $I^{2+}_{0}$
are singular in this limit we will rescale them and consider
integrals $a_n^{(2)}I^{2-}_{-4}$ and $a_1^{(1)} I^{2+}_{0} $
instead. In the result we obtain the following hamiltonians:
\begin{equation}
 I^{2-'}_{-4}(L(\l))\equiv\lim\limits_{a_n^{(2)}\rightarrow
 0}a_n^{(2)}I^{2-}_{-4}=1/2
\sum\limits_{i<n}\dfrac{(l^{(0)}_{in})^2}{a_i^{(2)}}, \ \ \
 I^{2+'}_{0}(L(\l))\equiv\lim\limits_{a_1^{(1)}\rightarrow
 0} a_1^{(1)} I^{2+}_{0}=1/2
\sum\limits_{i>1}\dfrac{(l^{(-1)}_{1i})^2}{a_i^{(1)}}.
\end{equation}
Their matrix gradients are written as follows:
\begin{equation}
 \nabla I^{2-'}_{-4}= \sum\limits_{i<n}
 \dfrac{l^{(0)}_{in}}{a^{(2)}_i}X_{in} ,
\  \
 \nabla I^{2+'}_{0}= \l^{-1}\sum\limits_{i>1}
 \dfrac{l^{(-1)}_{1i}}{a^{(1)}_i}X_{1i}.
\end{equation}
These are exactly $U$-operators of two independent generalized
Landau-Lifshiz hierarchies. That is why we call this hierarchy to
be { \it "doubled'' generalized Landau-Lifshiz hierarchy}.

Corresponding zero-curvature condition:
\begin{equation} \dfrac{\partial \nabla
I^{2-'}_{-4}}{\partial x_-}-\dfrac{\partial \nabla
I^{2+'}_{0}}{\partial x_+}+[\nabla I^{2-'}_{-4},\nabla
I^{2+'}_{0}]_{\mathcal{A}(\l)}=0.
\end{equation}
yields the following equations:
\begin{gather}\label{f1}
\partial_{x_+}l_{1i}^{(-1)}=-({a^{(1)}_i}/{a^{(2)}_i})l^{(-1)}_{1n}l^{(0)}_{in},\
\ \partial_{x_-}l_{in}^{(0)}=- a^{(2)}_i/a^{(1)}_il^{(0)}_{1n}l^{(-1)}_{1i},\\
\label{s1}
\partial_{x_+}l_{1n}^{(-1)}=a^{(1)}_n\sum\limits_{k=2}^{n-1}
l^{(-1)}_{1k}l^{(0)}_{kn}/{a^{(2)}_k}, \ \
\partial_{x_-}l_{1n}^{(0)}=
a^{(2)}_1\sum\limits_{k=2}^{n-1}l^{(-1)}_{1k}l^{(0)}_{kn}/a^{(1)}_k.
\end{gather}
Taking into account that functions  $$I^{2+'}_{0}(L^+(\l))= 1/2
\sum\limits_{i>1}\dfrac{(l^{(-1)}_{1i})^2}{a^{(1)}_i}=c_{-},\
I^{2-'}_{-4}(L(\l))= 1/2
\sum\limits_{i=1}^{n-1}\dfrac{(l^{(0)}_{in})^2}{a^{(2)}_i}=c_{+}$$
are constant along all time  flows, we obtain that
$l_{1n}^{(-1)}$, $l_{1n}^{(0)}$ are expressed via $l^{(-1)}_{1i}$
and $l^{(0)}_{in}$:
\begin{gather}
l_{1n}^{(-1)}=(a_n^{(1)})^{1/2}\bigl(c_{-}-
\sum\limits_{i=2}^{n-1}\dfrac{(l^{(-1)}_{1i})^2}{a^{(1)}_i}\bigr)^{1/2},
\ \ l_{1n}^{(0)}=(a_1^{(2)})^{1/2}\bigl(c_{+}-
\sum\limits_{i=2}^{n-1}\dfrac{(l^{(0)}_{in})^2}{a^{(2)}_i}\bigr)^{1/2}
\end{gather}
 and equations  (\ref{s1}) follows from the equations  (\ref{f1}).

Introducing for convenience   the following $(n-2)$
  component vectors: $$s_{-}^i=\dfrac{l^{(-1)}_{in}}{(a^{(1)}_i)^{1/2}},\ \ s_{+}^i=\dfrac{l^{(0)}_{in}}{(a^{(2)}_i)^{1/2}}, i\in 2,n-1.$$
and rescaling variables $x_{\pm}$  we obtain that our equations
acquire the following form:
\begin{gather}\label{fin1}
\partial_{x_+}\overrightarrow{s}_-=\Bigl(c_--(\overrightarrow{s}_-,\overrightarrow{s}_-)\Bigr)^{1/2}
\widehat{J}^{1/2}\overrightarrow{s}_+ ,\\ \label{fin2}
\partial_{x_-}\overrightarrow{s}_+=
\Bigl(c_+ -(\overrightarrow{s}_+,\overrightarrow{s}_+)\Bigr)^{1/2}
\widehat{J}^{-1/2}\overrightarrow{s}_-,
\end{gather}
where $(n-2)\times (n-2)$ matrix $\widehat{J}$ is defined as
follows: $\widehat{J}=diag((a_2^{(2)})^{-1}a^{(1)}_2,...
,(a^{(2)}_{n-1})^{-1}a^{(1)}_{n-1}).$

 {\bf Remark 8.} Note, that variables $\overrightarrow{s}_-$ could be
 expressed via $\overrightarrow{s}_+$ and its derivatives with
 respect to "negative time'' $x_-$ using equation (\ref{fin2}).
In the result one obtains system of nonlinear differential
equations of the second order on the vector
$\overrightarrow{s}_+$. Such procedure breaks simple
 form of the obtained equations and we prefer to leave
 them in the form of the system (\ref{fin1}-\ref{fin2}).

Let us now consider small $n$ example  of equations
(\ref{fin1})-(\ref{fin2}):

{\bf Example 1.} Let $n=3$. In this case  we obtain the following
two equations:
\begin{gather*}
\partial_{x_+}s_-=
\Bigl(c_--s_-^2\Bigr)^{1/2} j^{1/2}s_+ ,\ \
\partial_{x_-}s_+=
\Bigl(c_+ - s_+^2\Bigr)^{1/2} j^{-1/2}s_-.
\end{gather*}
Making substitution of variables: $s_{\pm}=c_{\pm}sin\phi_{\pm}$,
 and rescaling  variables $x_{\pm}$ we
obtain:
\begin{gather*}
\partial_{x_+}\phi_-=
\mathrm{sin} \phi_+ ,\ \
\partial_{x_-}\phi_+=
\mathrm{sin} \phi_-.
\end{gather*}
Expressing $\phi_-$ via  $\phi_+$ and putting it into the first
equation we finally obtain:
\begin{gather}\label{msg}
\partial_{x_+}\partial_{x_-}\phi_+=( 1- (\partial_{x_-}\phi_+)^2)^{1/2}
\mathrm{sin} \phi_+.
\end{gather}
This is exactly so-called "modified Sine-Gordon equation"
discovered by M. Kruskal and re-discovered later by H.Chen
\cite{Chen}( see also \cite{BZ} and references therein).

{\bf Example 2.} Let $n=4$. In this case equations
(\ref{fin1}-\ref{fin2}) define "the first negative flow'' to
standard Landau-Lifshitz equations. They have the  following form:
\begin{gather*}
\partial_{x_+}{s^1}_-=\Bigl(c_--((s_-^1)^2 + (s_-^2)^2)\Bigr)^{1/2}
{J}_1^{1/2}{s^1}_+ ,\ \
\partial_{x_-}{s^1}_+=
\Bigl(c_+ -((s_+^1)^2 + (s_+^2)^2)\Bigr)^{1/2}
{J}_1^{-1/2}{s^1}_-,\\
\partial_{x_+}{s^2}_-=\Bigl(c_--((s_-^1)^2 + (s_-^2)^2)\Bigr)^{1/2}
{J}_2^{1/2}{s^2}_+ ,\ \
\partial_{x_-}{s^2}_+=
\Bigl(c_+ -((s_+^1)^2 + (s_+^2)^2)\Bigr)^{1/2}
{J}_2^{-1/2}{s^2}_-.
\end{gather*}
Adding in  third component to the vectors
$\overrightarrow{s}_{\pm}$: $s^3_{\pm}=\Bigl(c_{\pm}
-((s_{\pm}^1)^2 + (s_{\pm}^2)^2)\Bigr)^{1/2}$, rescaling one of
the "times''  $x_-'= (J_1J_2)^{-1/2}x_-$ and making the following
change of indices in vector $\overrightarrow{s}_{-}$:
$s^2_{-}\longleftrightarrow s^1_{-}$ we obtain that the above
equations  are written as follows:
\begin{gather}\label{fin4}
\partial_{x_+}{s^1}_-= {s^3}_-
{J}_2^{1/2}{s^2}_+ ,\ \
\partial_{x_-'}{s^1}_+=
{s^3}_+
{J}_2^{1/2}{s^2}_-,\\
\partial_{x_+}{s^2}_-={s^3}_-
{J}_1^{1/2}{s^1}_+ ,\ \
\partial_{x_-'}{s^2}_+=
{s^3}_+ {J}_1^{1/2}{s^1}_-,\\
\partial_{x_+}{s^3}_-=-(
{J}_2^{1/2}{s^1}_-{s^2}_+  + {J}_1^{1/2}{s^2}_-{s^1}_+  ),\ \ \
\partial_{x_-'}{s^3}_+=-(
{J}_2^{1/2}{s^1}_+{s^2}_-  + {J}_1^{1/2}{s^2}_+{s^1}_-  ).
\end{gather}
This system of equations coincide with anisotropic chiral field
equations of Cherednik \cite{Cher}:
\begin{gather*}
\dfrac{\partial\overrightarrow{ s_-}}{\partial x_+}=[
\overrightarrow{s_-} \times
\widetilde{J}(\overrightarrow{s_+})],\ \ \
 \dfrac{\partial \overrightarrow{s_+}}{\partial x_-' }=[ \overrightarrow{s_+}
  \times \widetilde{J}(\overrightarrow{s_-})],
\end{gather*}
where a diagonal matrix $\widetilde{J}$ is defined as follows:
$\widetilde{J}=diag(J_1^{1/2},-J_2^{1/2},0)$.

\paragraph{Acknowledgements} I am grateful to P. Holod  for attention to the work
and to M.Pavlov for the discussion of small $n$ examples. The
research described in this publication was made possible in part
by INTAS Young Scientist Fellowship Nr 03-55-2233.


\begin{thebibliography}{99}
\bibitem{TF} Tahtadjan L. and Faddejev L. 1987 {\it Hamiltonian
approach in the theory of solitons} (Berlin: Springer) 586 p
\bibitem{New} Newell A. 1985 {\it Solitons in mathematics and physics}
(University of Arizona: Society for industrial and Applied
mathematics) 326 p
\bibitem{ZahSh} Zaharov V, Shabat A 1979 {\it Funct An  and
appl} {\bf 13} No 3 13-21
\bibitem{Kos}Kostant B 1979 {\it Adv Math} {\bf 34} 195-338
\bibitem{RST1} Reyman A, Semenov-Tian-Shansky M 1979 {\it
Invent. math.} {\bf 54} 81-100
\bibitem{RST3} Reyman A, Semenov-Tian-Shansky M 1989
{\it VINITI:  Fundamental trends} {\bf  6}, 145-147
\bibitem{Hol} Holod P 1984 {\it Proceedings of the international
conference" Nonlinear and turbulent Process in Physics'', Kiev
1983} (Harwood Head. Publisher) {\bf 3} 1361-1367
\bibitem{Hol1} Holod P 1987 {\it
Theoret and Math. Phys} { \bf 70} 11-19.
\bibitem{Hol3} Holod P 1987 {\it Soviet Phys Doklady} {\bf 32} 107-109
\bibitem{CP} Cantor I, Persits D 1988 {\it Proceedinds of the IX USSR conference in Geometry,
Kishinev, Shtinitsa} p.141
\bibitem{Bols}  Bolsinov A. 1988 {\it Trudy seminara po tenz. i vect. analizu}
 {\bf 23}, 18-28.
\bibitem{GS}   Golubchik I, Sokolov V  2000 {\it Theoret and Math.  Phys},
{\bf 124} No 1 62-71
\bibitem{Skr1} Holod P, Skrypnyk T 2000 {\it
Naukovi Zapysky NAUKMA} ser phys-math sciences {\bf 18} 20-25
\bibitem{Skr2}  Skrypnyk T  2001 {\it J Math Phys} {\bf 48} No 9
 4570-4582
\bibitem{Skr3}  Skrypnyk T, Holod P   2001
{\it J  Phys A: Mathematics and General} {\bf 34} No 9 1123-1137
\bibitem{Skr4} Skrypnyk T  2002 {\it  Czech J Phys}  {\bf 52 } No 11
1283-1288
\bibitem{Skr5} Skrypnyk T 2003 {\it Czech J Phys }  {\bf 53 }
No 11 1119-1124
\bibitem{Skr6}  Skrypnyk T.V. {\it J Math Phys} - to appear.
\bibitem{Chen} Chen H 1974 {\it Phys Rev Lett} {\bf 33}
No 15 925-930.
\bibitem{BZ}Borisov A,  Zykov S 1998
{Theor and Math. Phys } {\bf 115} No 2 199-214.
\bibitem{Cher} I.V. Cherednik  {\it Yadernaja Physica }{\bf 33} (1981) No 1, 278
\end{thebibliography}
\end{document}